FUNDP
NAMUR

# *If Entry Strategy and Money goTogether,*

# *What is the Right Side of the Coin?*

*by Annick Castiaux and Jean-Philippe Timsit*

Namur Center for Complex Systems

*Namur Center for Complex Systems*

*University of Namur*

*8, Rampart de la Vierge, B5000 Namur (Belgium)*

*http: // [www.naxys.be](http://www.naxys.be)*

# If Entry Strategy and Money go Together, What is the Right Side of the Coin?

**Jean-Philippe TIMSIT**
CRP Henri Tudor
29, avenue John F. Kennedy
L-1855 Luxembourg-Kirchberg
jean-philippe.timsit@tudor.lu
Tél.: +352 42 59 91 – 1
Fax: +352 42 59 91 - 777

**Annick CASTIAUX**
Louvain School of Management
Université de Namur (FUNDP)
Rempart de la Vierge 8 B-5000 Namur, Belgique
annick.castiaux@fundp.ac.be
Tel. +32 (0)81 72 48 80
Fax +32 (0)81 72 48 40

**Abstract:**

The goal of this study is to determine which strategic model, either IO or RBV, allows firms to generate the highest performance on a competitive market. Contrasting with classical studies that mobilize analyses as VARCOMP, we deploy a multi-agent system simulating the behavior of firms adopting RBV or IO strategic models. For an equivalent proportion of both strategic orientations, we study the instant and total performances of the firms on hyper-competitive markets. We show that the performance of best-performing IO firms, measured by the ROA, is higher in the short term, but that RBV firms obtain an average higher sustained performance, in the long term. Moreover, when they are in competition with IO firms on a highly profitable and competitive market, RBV firms which dare to enter such markets obtained generally the highest performance.

**Keywords**:

IO, RBV, Performance, Multi-Agent Simulation

## 1. INTRODUCTION

If, for the firm, resources and products are two faces of the same coin (Wernerfelt, 1984: 171), they are however not equivalent in term of performance. The question of the decomposition of performance is a key issue debated in the community of researchers in strategic management since many years. Indeed, the main function of firms is to generate performance, via the production of goods and services, to pay their owners and remunerate their employees (Durand, 2000). This question of performance, its underlying mechanisms and the appropriate level of analysis, fascinated economists and strategic management researchers since the 30s (Rumelt, 1974).

However, the mechanisms underlying the generation of this performance are particularly complex, which explains that they are the subject of a considerable literature of several thousand contributions (Capon, Farley, & Hoenig, 1990; Durand, 2000). Despite intense research, we still know little about the main sources of firm performance, and many questions remain unanswered.

Since twenty-five years and the pioneering work of Schmalensee (1985), a debate highly animates the community of researchers in strategic management (Arend, 2009). Is the performance of the firm mainly explained by the market or by the firm, is it fundamentally external or internal? Many studies suggest analyzing the components of firm performance on these two main levels of analysis (Hawawini, Subramanian, & Verdin, 2003; McGahan & Porter, 1997; Powell, 1996; Roquebert, Phillips, & Westfall, 1996; Rumelt, 1991; Schmalensee, 1985; Wernerfelt & Montgomery, 1988), most of them aiming to determine the main origin of the firm performance: industry, Strategic Business Unit (SBU) or the firm.

This debate carries with it the roots of a considerable challenge (Durand, 2000: 139-149). Indeed, if the main sources of firm performance are the market, it would demonstrate that the Industrial Organization approach (IO), based on the classical paradigm Structure - Conduct - Performance (SCP ) (Bain, 1951, 1968; Mason, 1939, 1957), best explains the performance of the firm, and a strategy based on this conception is potentially more profitable (Porter, 1980a; Porter, 1980b, 1991, 1996). If, however, the firm level explains better the performance generated by the firm, it would show the superiority of the Resource-Based View approach (RBV) (Amit & Schoemaker, 1993; Barney, 1991; Barney, 1986; Penrose, 1959; Peteraf, 1993; Wernerfelt, 1984) to explain the performance of the firm. So what is the strategic model which conducts the firms to generate more performance in a competitive market?

The divergent conclusions of previous studies (McGahan et al. 1997; Roquebert et al. 1996; Rumelt, 1991; Schmalensee, 1985) are hardly comparable because of the variety of methods of analysis of variance, numbers of years covered and sample sizes. These variations between the different studies do not allow the comparison between the empirical results, and do not guarantee the replicability to decide the question of the various components of performance, and thus the superiority of a possible model for other (Arend, 2009).

To cope with this difficulty of collecting reliable empirical data to determine the comparative advantages of the IO and RBV strategic models, we propose to reverse the question. The objective of this research is to use a technique of multi-agent simulation (Gilbert & Conte, 1994; Gilbert & Troitzsch, 2005; Harrison, Zhiang, Carroll, & Carley, 2007; Zott, 2003) to artificially produce data, over long periods, allowing flexibility and replicability of the analysis (Becker, Niehaves, & Klose, 2005; Moss & Edmonds, 2005). The model underlying our simulation considers two types of agents, in equal proportions, representing firms that operate in an environment of Pure Competition (PC). The first type of agents adopts decision rules corresponding to the IO strategic model, while the second type takes its strategic decisions according to RBV. Firms are acting in different market sizes with various expected return, then we look after a large number of cycles what is the strategy that has generated the highest performance. We led 1008 simulations, each including a series of 200 cycles (or periods). In each simulation, 200 firms and 20 markets interact.

The structure of this paper is as follows. We first present the theoretical background of this research (2), and the two strategic models, IO and RBV (3). We then detail the methodology used and the variable component model (4), then we present and discuss the results (5, 6), before concluding (7).

## 2. THEORETICAL CONTEXT

The question of the appropriate level of analysis of firm performance is an important issue that concerns the community of researchers in strategic management for many years (Table 1). It took a new turn since the pioneering contribution of Schmalensee (1985). Indeed, Schmalensee (1985) shows that the industry explains 19.5% of the variance in firm performance and that the share accounted for by the firm level is not significant. In view of this contribution, the performance of the firm would be mainly influenced by the market, which confirms the appropriateness of the IO model. In response to this study, conducted over one year and a very major error rate of nearly 80%, Rumelt (1991) replicated the analysis on the same database (FTC), but with four consecutive years. He showed (Rumelt, 1991: sample B) that industry accounts for only 4.03% of the variance in performance at firm level, instead of 19.5% in Schmalensee (1985). Besides this initial divergence, the share accounted for by performance levels and SBU firm, non-significant in the study by Schmalensee (1985), becomes significant in the study by Rumelt (1991): 1.64% of the variance explained by the firm and 44.17% of variance explained by the SBU. The proportion of variance explained by the same firm exceeds the share accounted for by industry, 17.9% against 10.2% in the study of Roquebert et al. (1996).

**Table 1: Comparison of the results of key studies, in percentage of explained variance**

| | % of the variance of the performance of firms by SBU explained by : | Schmalensee (1985) | Wernerfelt et Montgomery (1988) | Rumelt (1991) Sample A[a] | Rumelt (1991) Sample B[b] | Roquebert *et* Al. (1996) | McGahan et Porter (1997) | Hawawini et Al (2003) |
|---|---|---|---|---|---|---|---|---|
| Specifications of the studies | **Databases/nb of firms/nb of SBU/nb of industries** | FTC/456 firms/1775 SBU/242 industries | FTC/247 firms | FTC/457 firms/1774 SBU/242 industries | Same, but with less SBU | COMPUSTAT/6873 firms/13398 SBU/942 industries | COMPUSTAT/7003 firms/12296 segments/628 industries | Stern Stewart Dataset/562 firms/55 industries |
| | **Methods** | OLS & VARCOMP | OLS | ANOVA & VARCOMP | ANOVA & VARCOMP | VARCOMP | ANOVA & VARCOMP | VARCOMP |
| | **Performance measure** | rate of return | Tobin's Q | rate of return | rate of return | ROA | ROA | ROA |
| | **Years/Durations of the studies** | 1975/1 | 1976/1 | 1974-1977/4 | 1974-1977/4 | 1985-1991/7 | 1981-1994/14 | 1987-1996/10 |
| Results | **Industry*year** | n.a. | n.a. | 7,84 | 5,38 | 2,3 | n.a. | 3,1 |
| | **Market share** | 0,6 | -0,18 | n.a. | n.a. | n.a. | 4,33 | n.a. |
| | **Industry** | 19,5 | 12,30 | 8,3 | 4,03 | 10,2 | 18,7 | 8,1 |
| | **Firm** | n.a. | 2,65 (« focus effect ») | 0,8 | 1,64 | 17,9 | 4,3 | 35,8 |
| | **SBU** | n.a. | n.a. | 46,37 | 44,17 | 37,1 | 31,7 (segments) | n.a. |
| | **Error** | 79,9 | - | 36,87 | 44,79 | 32 | 48,4 | 52,0 |
| | **Total** | 100,0 | - | 100,0 | 100,0 | 99,5 | 100 | 98,0 |

**a** : Rumelt sample, very close from Schmalensee's (1985)
**b** : Rumelt sample, larger than Schmalensee's (1985)

The synthesis of key contributions (Table 1) highlights two key points. Firstly, the analysis of various studies presented Table 1 shows that firm performance attributable to the market level never exceeds 20%. It varies from 4.03% (Rumelt, 1991) to 19.5% (Schmalensee, 1985), which is consistent with Powell (1996). This result may seem surprising because a priori rather low compared to the share performance that the market does not explain, averaging over 80%. How to explain that such a small share of performance be explained? Thus, if the effects of the industry (Bain, 1951, 1968; Mason, 1939, 1957; Porter, 1980a; Porter, 1979, 1980b, 1996, 2008) cannot explain most of the variance in performance, what other factors can do it?

The second striking point is the difficult comparison between most of these works. Indeed, first of all, the methods are comparable, but different. The standard error is large, the smoothness of the method can improve the significance levels with a direct effect on results. Moreover, as levels of analysis, the durations of the studies vary. But the structure of a market varies with the intensity of competition (Mason, 1957). The period of analysis is vital. Finally, sample sizes vary as do sectors, and it is difficult to compare shares variance between several sectors where growth rates vary greatly from one sector to another. Therefore, the comparison between these studies does not guarantee the precision and replicability of thumb for deciding on the question of the different components of performance.

In the literature, the logic of the approach is still conducted through similar analysis of variance components (VARCOMP), in the light of that used by Schmalensee (1985). The analysis usually focuses on the overall performance of many multi-product firms, a large number of industries, determining how the variance of performance breaks down market, firm and SBU. This logic has shown, in our view, its limitations. Performance is measured mostly by the accounting measure of profitability, that is to say the “Return On Assets” (ROA), with the exception of Montgomery and Wernerfelt (1988) who use Tobin’s Q.

The best would be to benefit from a comparison between the strategy adopted by a firm at t time, and its performance after several periods. However, the problem is twofold. Firstly, it is complex to apply to decision makers all the information about the strategy they have decided to run for years to come, and then compare it to as the years passed with the performance occurred. Indeed, this information is vital to the firm and is therefore not transmitted. In addition, the firm undergoes external changes, especially regarding target markets, and internal, for example in management teams, which influence the strategic concepts. Strategies thus undergo inflections that must be taken into account. Alternatively, ask a manager to explain the choices made in the past gives a result very questionable in light of the effects of post hoc rationalization.

The two main slopes of strategic management are rivals to explain the performance and growth of the firm. To untangle this hank, it is necessary to decompose and compare the two different generation mechanisms of performance, according to the two strategic models studied.

## 3. TWO RIVAL CONCEPTIONS

When Wernerfelt chose purposely the first sentence of his seminal article "For the firm, resources and products are two sides of the same coin" (Wernerfelt, 1984), he highlighted the two opposing dimensions of strategic management, still young science at the time. By "resources", he means the internal conception of the strategic modeling, based on an analysis of the components of the firm; and "products ", an external conception, based on an analysis of the environment.

These two strategic models are not homogeneous but both fragmented, based on strong conceptual foundations. The IO approach is a dome theory covering five separate streams (Conner, 1991b): the neoclassical theory's perfect competition model (Alchian & Demsetz, 1972), the Bain and Mason model (Bain, 1951, 1968; Mason, 1939, 1957), and the responses of Schumpeter (Schumpeter, 1951) and the School of Chicago , and finally the transaction costs theory (Williamson, 1975, 1985; Williamson, 1991). To these streams, we can add marginally the contestable markets theory (Baumol, 1982; Baumol & Willig, 1981). The foundations of the IO model are based on the paradigm Structure-Conduct-Performance, initiated by Mason (Mason, 1939, 1957) and Bain (Bain, 1951, 1968), and then extended and generalized by Porter (Porter, 1980a; Porter, 1979, 1991, 1996), for example through the analysis of competitive forces. This model is based on the fact that market structure determines the behavior of firms and thus their performance. Strategic modeling is initiated by an external analysis to the firm.

The resource-based theory takes its origins in the work of Penrose (Penrose, 1955; Penrose, 1952; Penrose, 1959), in reaction to the neoclassical conception of the firm as a black box, and the model of pure competition, one of the foundation of IO. It has been later popularized by Wernerfelt (1984), but especially by Barney (1986, 1991). According to RBV, the firm is composed of resources and distinctive capabilities that make it unique in its market. In this conception, the growth of the firm takes mainly two paths. The first one is the accumulation and growth by deploying resources. The second is the creation of value by combining resources, thanks to the skills and abilities held by the firm (Amit *et al.*, 1993; Barney, 1991; Barney, 1986, 1996; Conner, 1991a; Teece, Pisano, & Shuen, 1997; Wernerfelt, 1984). However, the resources of the firm are indivisible (Penrose, 1959: 67-71) and are not homogeneous as factors of production can be. This conception of the firm also gave rise to several approaches like Knowledge-Based View (KBV) (Foss & Knudsen, 2003; Grant, 1996; Spender, 1996; Spender & Grant, 1996; Tsoukas, 1996), Competence-Based View (CBC) (Hamel, 1991; Hamel & Heene, 1994; Prahalad & Hamel, 1990), or Dynamic capabilities (Eisenhardt & Martin, 2000; Helfat, Finkelstein, & Mitchell, 2007; Teece, 2007; Teece *et al.*, 1997), to name a few.

This study focuses on the fundamentals of these two currents. On the IO approach, we focus on the classical origins of this strategic model, namely the work of Mason (1939, 1957), Bain (1951, 1968) and Porter (1979, 1980a, 1980b, 1991, 1996). For the RBV, we focus again on its conceptual foundations, principally on the contributions of Penrose (1959), Wernerfelt (1984, 1994) and Barney (1986, 1991).

This study, we focus specifically on two key points: the level of analysis and mechanisms of performance generation across the sources of competitive advantage.

**Level of analysis**

Originally, Mason (1939, 1957) raises the market as the relevant level to analyze the performance of firms, a model subsequently developed and deepened by Bain (1951, 1954) and popularized by Porter (1979, 1980, 1981). This design lays the foundation of the industrial organization approach (IO) arguing that the market structure, measured primarily by the concentration of producers, the concentration of buyers, the degree of differentiation between products and the conditions of entry for new competitors (Durand, 1997, 2000;

Scherer & Ross, 1990), determines the behavior of players. This behavior has the direct consequence of firm performance. Based on this view, Porter (1979, 1980, 1981) provides an analysis of market structure based on five competitive forces (Porter, 1991, 1996, 2008).

Unlike the IO model, the resources conception conceives the firm as the relevant level of analysis. Indeed, Penrose (1959) presents the firm as a collection of resources and skills. This design is a reaction to the neoclassical approach that models the behavior of agents in a market by reducing the firm to a black box modeled by a production function Q = f (K, L). Thus, as the RBV model (Amit *et al.*, 1993; Barney, 1991; Barney, 1997; Barney, 1986; Dierickx & Cool, 1989; Grant, 1991; Penrose, 1959; Peteraf, 1993; Wernerfelt, 1984), the appropriate level of strategic analysis is the firm because it is combining its resources the firm product performance (Peteraf, 1993; Teece *et al.*, 1997).

**Mechanisms of performance generation**

The question of the sources of the competitive advantage is an important point of opposition between the IO and RBV models. The I.O. analysis conceives that the market can produce a total amount of profit. This total profit is shared among all firms in the market. The goal of every firm is to access a share of profit above the average profit by acting on its suppliers and customers, but also on its competitors, and thereby altering the market structure to improve its position on it. Through this, the objective is being to reach a situation where monopoly rents are maximized (Porter, 1991, 1996; Ricardo, 1817). The greater the number of competitors is reduced, the larger the share of profit on the remaining firms increases. A rent, as an extra profit, is generated by the increased market power. This rent is the extra profit which firms in a market can access, due to the specific configuration of market monopoly or duopoly example (Mahoney & Pandian, 1992; Ricardo, 1817). However, the larger the share of profit by business grows, it makes the market more attractive, attracting companies from other markets (Baumol, 1982; Baumol *et al.*, 1981). If these firms enter the market, they will lower the average profit share. The goal of firms is to reduce the number of competitors in the market, while protecting themselves by barriers to market entry, thus discouraging other firms from entering.

Alternatively, the RBV approach conceives the firm as level of analysis, it is at this level that plays the issue of competitive advantage. The firm is composed of resources that constitute its substance (Penrose, 1959; Wernerfelt, 1984. These resources are combined using the skills and capabilities to produce goods and services to the market (Amit *et al.*, 1993; Barney, 1991; Dierickx *et al.*, 1989; Durand, 1997; Grant, 1991; Hamel *et al.*, 1994; Makadok, 2001; Peteraf & Barney, 2003; Teece *et al.*, 1997). Some of the resources that make up the firm have particular characteristics (Barney, 1991; Barney, 1997): they are valuable, rare, inimitable and non-substitutable (VRIN). The source of competitive advantage is in the possession of these resources VRIN (Barney, 1997)and in the ability of the firm to implement them, that is to say, to combine and deploy them in mobilizing routines more or less refined (Teece, 2007; Teece *et al.*, 1997; Zollo & Winter, 2002).

For Barney and the main theorists of RBV, the analysis of competitive market forces do not give enough information to the firms to build a competitive advantage (Barney, 1986). The superiority of the RBV model on the IO conception lies in the analysis of resources and skills that form the firm (Barney, 1986: 1234). Analysis of Barney specifically stresses that strategic choices are mainly derived from the analysis of skills and unique capabilities of the firm rather than the environment.

IO and RBV approaches on levels of analysis and design of competitive advantage are very different, so the consequences on the formulation of the strategy are very important. These differences have a significant impact on many segments of strategic thinking such as the structure of the firm, the diagnosis of competitive forces, or the value creation mechanisms. However, the point at which this difference is in our opinion the most glaring is the implementation of the strategy for market entry. According to pursue that strategy is based primarily on a market analysis or an analysis of the firm, reasoning led to produce the performance will be drastically different, even opposite (Teece *et al.*, 1997). This question of formulating the entry strategy is in our opinion the most striking feature between the two behaviors, it is here that we will focus the simulation.

## 4. METHODS

Our goal is to understand the individual behaviors of and interactions between firms, markets and resources; in this framework, the ideal method is the agent-based simulation technique (Becker *et al.*, 2005; Demazeau & Müller, 1990). This method is based on the creation of different agents following behavior and interaction rules and evolving in a common environment. In our case, agents are firms and markets on which firms can evolve.

Building the simulation requires a preliminary modeling of key agents (firms, markets) as well as the determination of the dependent variable, firms' performance, and the independent variables. The analyzed firms evolve in an environment made of markets on which they commercialize goods and services in order to generate performance. To sustain their production activity, they must obtain assets in order to maintain their activity. Thus, the simulation has to take into account the modeling of the environment components (a market of goods and services) and of the behavior of firms in relationship with the market of strategic factors (Barney, 1986).

To simplify the model and to facilitate resource representation, we reduced their number in comparison with the usual typology of resources as found in the literature (Barney, 1986; Grant, 1991). We opted for a financial resource as well as three other kinds of resources, which could be organizational, human or physical resources. We chose to attribute a color to those three resource types in order to represent them visually by the three fundamental colors: red, green and blue. As a matter of fact, numerical simulations provide the opportunity to generate visual representation of agents' evolution. We created a Youtube account where readers can visualize several simulations performed for this research (see http://www.youtube.com/watch?v=Z4HjrLsBRXQ). [Lien non fonctionel pour le moment]

To summaries, the key elements of the model are thus firms' performance, markets characteristics, firms' behaviors and strategic factors market. We explain each of those elements in detail in the following subsections.

**Firms' performance**

The question of performance and its multiple measures was remarkably treated in the classical paper of Capon *et* al. (1990). Generally, in strategy literature, the measure of performance is the economic profitability modeled by the ROA (“*Return on assets*”) (Lee & Madhavan, 2010), which leads to reduce the total financial performance of the firm by the total value of its assets, in order to determine the produced wealth by asset unit. We align ourselves on this common practice, using the ROA to measure our firms' performance, which also helps us to control size effects (as amounts of assets) between firms. The ROA is computed as the ratio between net revenues and assets. We compute each firm's profit as follows:

$$\boldsymbol{\Pi = TR - TC}$$

where:

- Π is the profit;
- TR is the total revenue computed as the total number of sold goods and services, multiplied by their price, i.e. P*Q;
- TC are the total costs supported by the firm, following production.

We divide the firm's profit by the value of its resource portfolio, obtaining then the profitability by unit of assets, which measures the ROA. However, as our simulation takes place on a large number of periods, we compute two measures of this performance: the instant performance $P_i(t_j)$ obtained at each period:

$$P_i(t_j) = ROA_i(t_j) = \frac{\Pi_i(t_j)}{RES_i(t_j)}$$

where $RES^i_t$ is the total value of the resources owned by firm i on time $t_j$.

We also compute a total performance, which is an actualized value of the ROA from the beginning of the simulation, $P^t_i(t_n)$:

$$P_i^T(t_n) = \sum_{j=1}^{n} ROA_i(t_j)$$

We have thus two indicators at our disposal: the instant performance of the firm, allowing, at each period, to evaluate firms' performance following their strategic choice, and the total performance of firms allowing to evaluate each strategic type for the duration of the simulation.

**Markets in Pure Competition**

Firms' performance is linked to numerous parameters (Milgrom & Roberts, 1995) and all firms interact in an environment made of numerous markets submitted to considerable forces. Among others, the location of confrontation of those forces, and thus the nature of the analyzed markets is fundamental. In order to simplify the model, we chose to simulate an environment made of markets respecting the conditions of Pure and Perfect Competition (PPC):

- **Atomicity**: All firms on a given market are « price taker » and we consider P as fixed to P=1; only Q is variable ;
- **Homogeneity**: all firms manufacture the same product on a given market;
- **Mobility of production factors**: Production factors move freely and there is a resource market called Market of Strategic Factors (SFM)(Barney, 1986);
- **Transparency**: information is perfect, without costs.

Nevertheless, the perfect information hypothesis raises several issues. Following path dependency (Arena & Lazaric, 2003; Nelson & Winter, 1973; Teece *et al.,* 1997; Winter, 1990), and thus firms' age, their analysis of market evolution and potential future value of resources (both inside the firm and on the SFM) evolves. We thus decided to relax the perfect information hypothesis introducing a random coefficient for two factors: the realization of an anticipated performance by a given firm and the evolution of resource prices, initially submitted to law of supply and demand. Relaxing this hypothesis has a strong impact on the way the SFM is working. As a matter of fact, firms do not acquire resources exclusively following their value on t, but also following their anticipated value on t+1.

**Firms' behavior**

As we said before, we analyze the strategic behavior of firms through their entry strategy on markets. A firm behaving following I.O. principles will formulate its entry strategy as follows (Teece *et al.*, 1997):

1- Choice of a market following its attractiveness (structure: number of competitors, reachable profit share, …);
2- Choice of an entry strategy based on the strategy of market competitors;
3- Acquisition of needed resources to be competitive on the market (if those resources are not yet in the firm's portfolio).

A firm behaving following RBV principles will formulate its entry strategy differently:

1- Identification of the specific resource providing itself with a competitive advantage;
2- Determination of markets where those resources can lead to the highest profits;
3- Evaluation of maximal profits generated by the strategic assets among the following possible actions:
    a. Market entry
    b. Selling of the resource on the SFM
    c. Selling of resource outputs

Where the IO firm initiates its strategy by a diagnostic of the competition field, the RBV firm begins with a diagnostic of its resources and routines.

**Strategic Factor Markets**

Firms can acquire or sell resources on the SFM (Barney, 1986). Following the relaxation of the perfect information hypothesis, the purchase cost of resources is fixed by the law of supply and demand, tempered with a partially random factor, linked to firms' age. As a matter of fact, regarding path dependency, firms' age conditions the constitution of resource collection and the formation of organizational competencies and capacity allowing the implementation of those resources (Arena *et al.*, 2003; Nelson *et al.*, 1973; Teece *et al.*, 1997; Winter, 1990).

The SFM is a market where resources required to implement a strategy are bought. It is composed of firms, themselves composed of resources. Those resources – the word "assets" is more relevant when they do not belong to a firm – are produced by institutions, as universities or schools (human resources). All strategies where the acquisition of resources is require need to call this SFM into play (Barney, 1986).

Firm's performance do not only depend on the compliance between the strategy and the market, but also on the strategy implementation costs. Firms estimate the future value of the acquired strategic resources following the resources they owe, their competencies and capacities, and the strategy they intend to implement. The resource purchase decision is thus not only based on an intrinsic cost, but also on an anticipated future value, and by the law of supply and demand.

If suppliers and demanders have the same vision of the future value of the resource, then the price of the resource will approximately correspond to its value. Otherwise, suppliers will sell the resource at a higher (respectively lower) price, and / or buyers will buy the resource above (respectively below) its value, or the resource will not find a buyer. Thus, some resources will be very costly because several firms anticipate a high future value creation, while other resources are under-evaluated because not so demanded in the implemented strategies.

However, the evaluation of resource future values is linked to PPC, which assumes perfect information: the same information for all actors, at the same time, and thus the same vision of future winning strategies. This is not true. Information being imperfect, the SFM includes a certain degree of uncertainty. For Barney (1986), one of the important success criteria is thus luck. Luck can provide an over-performance through the lucky acquisition of the winning resources on the SFM.

To summarize, the value of a resource on a SFM is linked to its cost and its value anticipated by the firms following the formulated strategies. If a resource s affordable, but numerous firms formulate strategies using this resource, anticipating an important profitability at term, then the resource value on the SFM will rise drastically. However, this hypothesis is linked to perfect information. We relaxed this hypothesis. Moreover, the impact of chance in the acquisition of resources and evolution of markets, allowing an unpredictable success or failure, has also to be taken into account.

## Technical characteristics of the simulations

The technical characteristics of the simulation are as follows. As modeling software, we used NetLogo version 4.1. We led 1008 simulations, each including a series of 200 cycles (or periods). In each simulation, 200 firms and 20 markets interact. Markets can have 3 different sizes, following the number of shares they include: 10, 100 or 1000. Those market shares represent the demand intensity for each market. Firms are represented by arrows which color is the combination of their amount of red, green and blue resources. Initially, firms receive a random combination of those resources, but this combination – and thus the firm's color – evolves through cycles. Markets are represented by circles with different diameters following their size and with different colors following the resources they require from firms to act efficiently – i.e. produce goods and services – on them.

A given firm I is characterized by (1) a strategic orientation IO or RBV, fixed through time and determining their entry modalities on one market or another; (2) financial resources, initially identical for all firms and evolving following resource purchases and, if any, cumulated profits; (3) resources of three types ($r_i(t)$, $g_i(t)$, $b_i(t)$), of which they receive a random initial stock and that they can buy or sell on the SFM following their needs, i.e. the requirements of the market on which they try a venture or the possible financial difficulties they meet. Each firm takes a different color, representing its resource collection, and reflecting its unique combination of red, green and blue resources; (4) an instant performance $P_i(t)$, measured by the ROA, corresponding to the profit realized on t, divided by the total value of the firm's resources at the same time; (5) a total performance $P^T{}_i(t)$, measured by the summation of instant ROA, corresponding to the sum of instant performances from the initial time.

A market j is characterized by (1) a fixed size $NP_j$ corresponding to the number of available market shares; (2) a value of market shares $v_j(t)$ varying through time following the concentration of firms on the market; (3) fixed access conditions, simulated by a minimal level of the three fundamental resources, $R_j$, $G_j$ et $B_j$, that firms must owe to enter this market.

Firms and markets interact, as shown on Figure 1, by the intermediary of the SFM, on which prices evolve following the law of supply (available stocks on the SFM and inside firms) and demand (access requirements by all markets), and of which stocks evolve following purchases and sales performed by firms to penetrate a market or restore their finances.

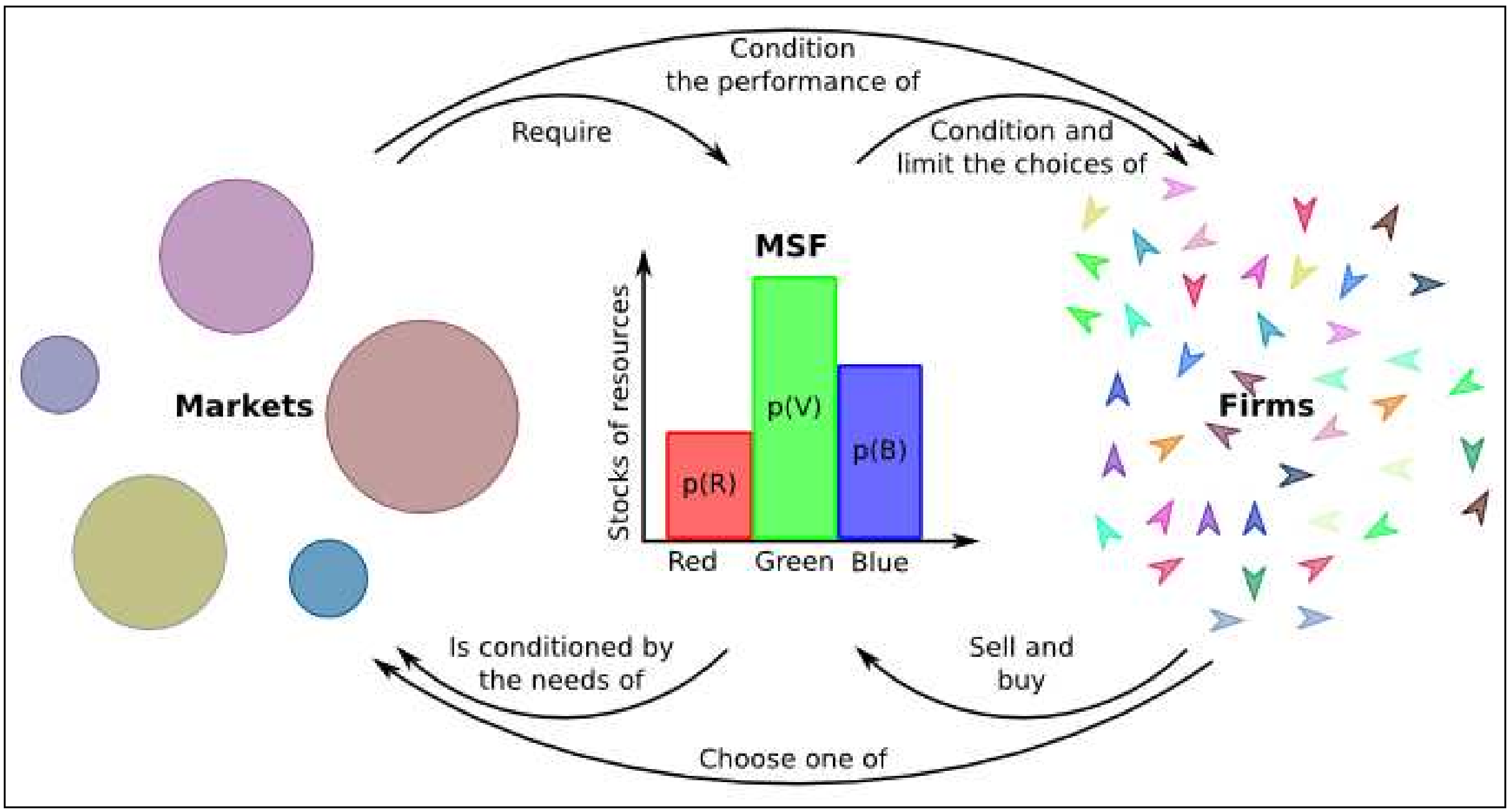

**Figure 1: Model components and mutual interactions**

On a given time t, according to whether the strategic orientation of the firm is IO or RBV, it will look for a market using different behavioral rules. If firm I is IO, it will search, among the M available markets, the one promising the highest profit, computing:

$$\underset{j=1,M}{Max}\, \Pi_i^j(t) = \underset{j=1,M}{Max} \frac{NP_j \times v_j(t-1)}{NF_j(t-1)}$$

If firm is RBV, it will look for the market for which its collection of resources provides the best asset to face the market entrance barriers (i.e. the required resources to enter the market). Thus, it will search, among the M available market, the one for which it already owes the necessary resources or, if no market meets this condition, the one minimizing the resource investment. We simulated this search by computing the Euclidian distance between the firm's resource portfolio and the market's entrance barrier, as follows:

$$\underset{j=1,M}{Min}\ d_i^j(t) = \underset{j=1,M}{Min}\ \sqrt{(pos(r_i(t)-R_j))^2 + (pos(g_i(t)-G_j))^2 + (pos(b_i(t)-B_j))^2}$$

where:

$$pos(x) = \begin{matrix} x\ si\ x \geq 0 \\ 0\ si\ x < 0 \end{matrix}$$

## 5. RESULTS

After conducting 1008 simulations of 200 cycles, we present the results in two tables. Table 2 summarizes the raw results of the simulations. This study simulates strategies in hypercompetitive markets, firms acting against each other. We therefore present table 3 which summarizes the simulation results in relative values. Each of these results is presented in two parts, period 20 and period 200, to highlight the direct effects of market entry, and the consequences of long-term strategies.

**Tableau 2: Synthesis of the raw results and comparison between cycle 20 and cycle 200, for the 1008 simulations**

| | Nb. IO in best 10 After 20 cycles | Nb. RBV in best 10 After 20 cycles | Nb. IO in best 10 After 200 cycles | Nb. RBV in best 10 After 200 cycles | Perf. 1st IO After 20 cycles | Perf. 1st IO After 200 cycles | Perf. 1st RBV After 20 cycles | Perf. 1st RBV After 200 cycles | Av. Perf. 10 Best IO After 20 cycles | Av. Perf. 10 Best IO After 200 cycles | Av. Perf. 10 Best RBV After 20 cycles | Av. Perf. 10 Best RBV After 200 cycles |
|---|---|---|---|---|---|---|---|---|---|---|---|---|
| **Average** | 5,55 | 4,45 | 3,93 | 6,07 | 3447,23 | 5582,62 | 3528,57 | 6654,33 | 3146,82 | 5339,84 | 3011,54 | 5585,26 |
| **St. dev.** | 2,71 | 2,71 | 3,27 | 3,27 | 253,39 | 294,05 | 458,88 | 937,24 | 193,73 | 292,25 | 320,26 | 616,07 |
| **Variance** | 7,36 | 7,36 | 10,68 | 10,68 | 64204,78 | 86465,27 | 210568,59 | 878420,96 | 37530,87 | 85412,89 | 102567,34 | 379540,78 |
| **Median** | 6,00 | 4,00 | 4,00 | 6,00 | 3444,99 | 5597,86 | 3483,70 | 6587,88 | 3156,50 | 5350,85 | 3052,31 | 5668,28 |
| **Maxima** | 10,00 | 10,00 | 10,00 | 10,00 | 4152,83 | 6432,88 | 4962,37 | 8490,24 | 3673,89 | 6098,50 | 3715,42 | 6812,13 |
| **Minima** | 0,00 | 0,00 | 0,00 | 0,00 | 2673,57 | 4535,61 | 1463,60 | 1827,73 | 2533,42 | 4298,64 | 1002,85 | 1560,73 |

**Table 3: Synthesis of the relative results, and comparison between cycle 20 and cycle 200, for the 1008 simulations**[1]

| | Rel diff between best IO and best RBV[a] After 20 cycles | Rel diff between best IO and best RBV[a] After 200 cycles | Rel diff between av 5 best IO and av 5 best RBV[b] After 20 cycles | Rel diff between av 5 best IO and av 5 best RBV[b] After 200 cycles | Rel diff between av 10 best IO and av 10 best RBV[c] After 20 cycles | Rel diff between av 10 best IO and av 10 best RBV[c] After 200 cycles | Rel diff av all IO and av all RBV[d] After 20 cycles | Rel diff av all IO and av all RBV[d] After 200 cycles |
|---|---|---|---|---|---|---|---|---|
| **Average** | -0,01 | -0,14 | 0,03 | -0,08 | 0,06 | -0,02 | 0,82 | 0,66 |
| **St. dev.** | 0,16 | 0,17 | 0,16 | 0,17 | 0,20 | 0,20 | 1,05 | 0,89 |
| **Variance** | 0,03 | 0,03 | 0,03 | 0,03 | 0,04 | 0,04 | 1,09 | 0,80 |
| **Median** | -0,02 | -0,15 | 0,00 | -0,10 | 0,03 | -0,05 | 0,66 | 0,51 |
| **Maxima** | 1,42 | 2,08 | 1,98 | 2,35 | 2,37 | 2,60 | 19,42 | 14,25 |
| **Minima** | -0,37 | -0,42 | -0,28 | -0,36 | -0,24 | -0,32 | -0,05 | -0,16 |
| **Nb of cases where IO > RBV** | 443,00 | 143,00 | 511,00 | 195,00 | 605,00 | 331,00 | 1007,00 | 994,00 |
| **Nb of cases where RBV > IO** | 565,00 | 865,00 | 497,00 | 813,00 | 403,00 | 677,00 | 1,00 | 14,00 |
| **% of cases where IO > RBV** | 43,95% | 14,19 % | 50,69% | 19,35 % | 60,02% | 32,84 % | 99,90% | 98,61 % |
| **% of cases where RBV > IO** | 56,05% | 85,81 % | 49,31% | 80,65 % | 39,98% | 67,16 % | 0,10% | 1,39 % |

Where :
**a** : Relative difference between the best performing IO and RBV firms : (Best IO – Best RBV)/Best RBV
**b** : Relative difference between the average performance of the 5 best performing IO and RBV firms: (Best5IO – Best5RBV) / Best5RBV
**c** : Relative difference between the average performance of 10 best performing IO and RBV firms : (Best10IO – Best10RBV) / Best10RBV
**d** : Relative difference between the average performance of all remaining IO and RBV firms : (AvIO – AvRBV) / AvRB

[1] See appendices 1 & 2 for full results

The first point to note in analyzing the raw results is the dramatic twist for the 10 best performing firms between period 20 and period 200. Indeed, averaged over the 1008 simulations, in period 20, there are 5.55 IO firms in the 10 best performing companies, against 4.45 RBV firms. This shows that IO firms access higher performance faster than RBV firms, on average. However, it seems that they do not succeed to maintain this performance. Indeed, at period 200, there are on average no more than 3.93 IO firms present in the top 10 best performing firms, against 6.07 RBV firms.

A second observation is that RBV firms succeed to grow more steeply during the period. Indeed, if we compare the performance (in value) of the top 10 firms in each strategic orientation between cycle 20 and cycle 200, IO firms present an average increase of 41% (from 3146.82 to 5339.84). In contrast, RBV firms demonstrate an average increase of 46%, i.e. nearly 5 points more than IO firms, from an average performance of 3011.54, lower than IO firms' average performance, to an average performance of 5585.26 higher than IO firms' average performance.

Table 3 analysis goes in the same direction. Indeed, if one examines the column summarizing what strategic model has been adopted by the most efficient firm, on 1008 simulations, it appears that, after 20 cycles, RBV leads with 56.05% of the cases against 43.95% of the cases for IO. Thus, even if RBV firms, has demonstrated table 2, are minority in the top 10 best performing firms, in more than half of the cases an RBV firm is the most efficient. This result strengthens in cycle 200. Indeed, at this stage, the highest performance, after 1008 simulations, is reached by an RBV firm in 85.81% of the cases. So, it seems that after 200 cycles, IO firms have lost the advantage they demonstrated after 20 cycles, and that the best performing RBV firms are clearly sustaining a competitive advantage that pays in time.

However, the column "Rel diff av all IO and av all RBV After 200 cycles" gives a result that might seem contradictory. Indeed, we see that after 200 cycles, in 98.61% of cases, the IO strategic model is binding on the RBV model (1.39%). These results show that firm performance varies widely from one RBV firm to another. If the low performing RBV firms recuperate a very small competitive advantage after 200 cycles, they are clearly underperforming in comparison with IO firms.

Indeed, while IO firms' behavior is fairly homogeneous, RBV firms' behavior seems much more heterogeneous. The analysis of the results of 1008 simulations allows us to identify three different types of behavior RBV firms. The first type of RBV firms includes all firms that have not been able to quickly get a profitable position on a market in line with their resources collection and abilities. In these cases, the performance of these firms is close to zero, or even negative. This has a strong impact on the average performance of all RBV firms. The second case concerns RBV firms that succeed to reach a market, but on which no IO firm decided to get positioned because of a too low expected gain which is too low. The performance of these firms is low, but they manage to perform enough to survive. At last, the third case concerns hypercompetitive markets. The RBV firms which go on hyper-competitive markets, where they seem to perform even better than their direct IO competitors. This is partly a question of chance because they have the resources collection and abilities to face this market. This will be still more obvious when looking at performance further in time. One can also consider that those firms are also risk-taking firms, daring to go on markets where competition is very intensive, but on which they benefit from competitive advantages, that IO firms have still to acquire and fail to maintain.

At this point, the constraints of the model should be taken into account in the analysis of the results. Firstly, a random collection of resources given initially to firms does not guarantee them to find a market with an RBV decision-making process. Instead, IO firms will always find a market that matches their criteria. Secondly, firms can not change their strategy, which means that RBV firms are locked once they have found a market, even if it is non profitable. Thirdly, firms can easily find resources. Actually, we introduced low constraints on resource stocks on the Strategic factor Market. This guarantees IO firms, if they have enough finances, to find the relevant resources for any market they would target. If a higher constraint on resource availability would be set, this would also eliminate IO firms from the picture, or at least make their life more difficult.

We present hereafter (Figure 2) a representative case, compliant with the results of our 1008 simulations. We compare the evolution of the two dependent variables – instant and total performances – all along the 200 simulation periods for the best 5 firms of each strategic orientation. Abscissa show time periods while ordinates must be read as monetary units. The difference between IO firms (above) and RBV firms (below) is significant for both variables.

During the first cycles, the best performing IO firms often demonstrate a higher instant performance than RBV firms, but this situation does not last. After this initial period, occasionally, the instant performance of one or another IO firm is exceptional, distancing the instant performances reached by RBV firms. Between those great feats, coming from a successful entry on a highly profitable market, IO firms generally demonstrate a low, if not negative, performance. RBV firms, for their part, demonstrate a more sustained performance, decreasing naturally with time, but maintained at levels allowing to reinforce continuously the position of these firms on the chosen market.

This sustainability of RBV firms' results gives a total performance growing regularly. On the other hand, the total performance of IO firms grows by fits and starts, under the impetus of a good instant performance. At the end of the simulation, the regular performance of RBV firms pays, since they reach a total performance clearly superior to the one reached by IO firms.

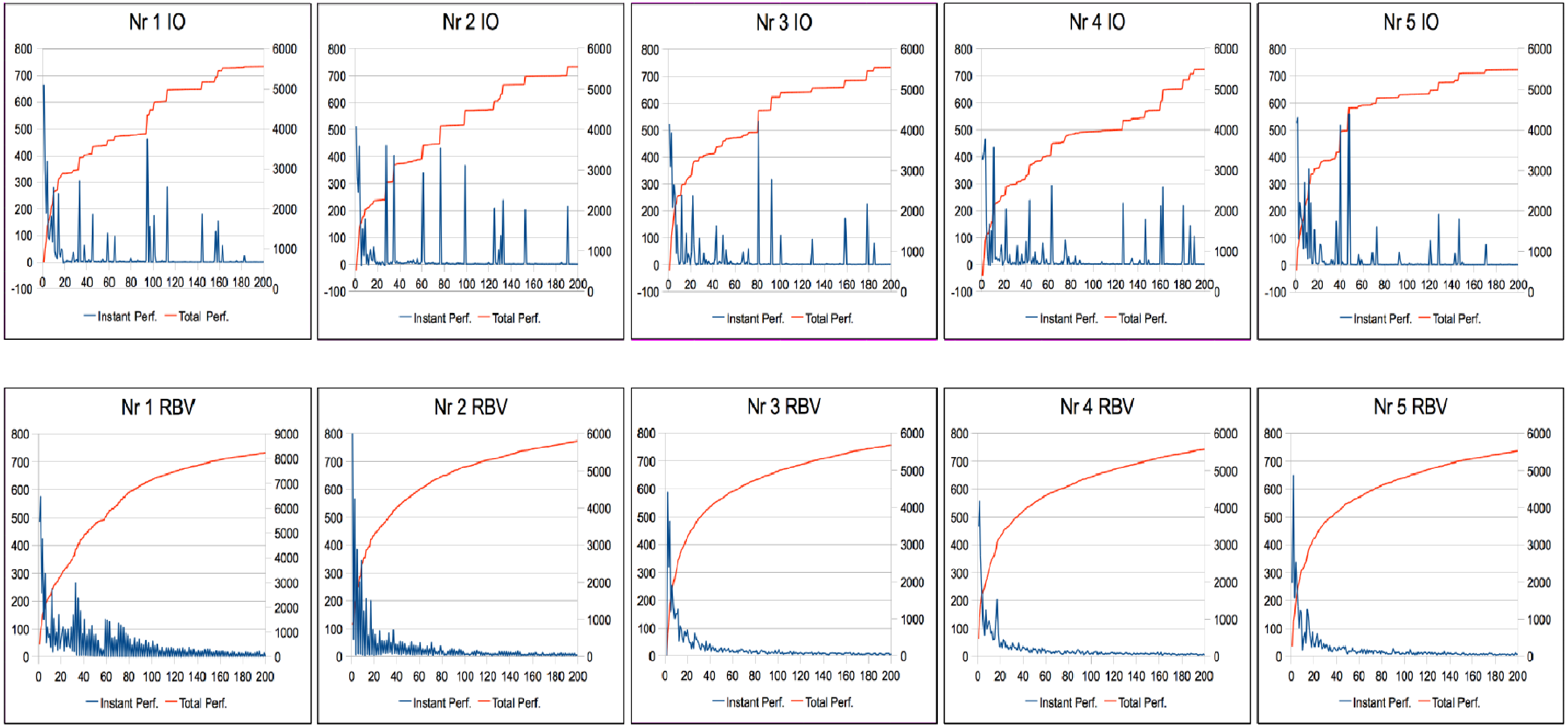

**Figure 2: Graphical representations of the instant and total performances of the 5 best IO (above) and RBV (below) firms, for representative simulation among the 1008 simulations**

## 6. DISCUSSION

Saying that the initial instant performance of IO firms is higher is tautological. As a matter of fact, the market choice algorithm for IO firms (Teece *et al.*, 1997) is based on market expected performance. IO firms target and enter markets promising a higher than average performance (Porter, 1991). It is thus normal that, in the short term, their performance is high. Nevertheless, the simulations show that IO firms are not good at perpetuating this over-performance. They reach very high performances but obtain also remarkably disappointing performances. When actualizing those performances on the total duration of the simulations, highest performances do not compensate lowest performances. This is obviously illustrated by the curves presenting total performances (Figure 2) of IO firms. Those curves present sharp steps, which show that growth phases are not continuous. On the contrary, IO firms undergo performance shocks acting as strong but temporary upsurges, not allowing to reach a final high total performance, the firm suffering from the low performance periods.

Inversely, RBV firms present lower instant performances, but realize, in average, less lower performances than IO firms; their total performance curve is thus continuous and is exempt from shocks.

This is partly explained by "luck and expectations" (Barney, 1986), because some RBV firms quickly found a market in which their collection of resources allows them to be successful. Thus, the speed with which a firm identifies the market appears suitable as a key to success (Teece, 2007). Moreover, when the market is composed of a large number of IO firms relative to RBV firms, RBV firms appear to have a performance bonus. The keys of this outperformance seem to be daring, speed of reaction and the competitive structure of the market through a significant proportion of IO firms. When these conditions are met, the RBV strategic model is clearly superior to the IO model.

On the entire simulations time, successful RBV firms succeed to reach a total performance higher than IO firms. This result would be consistent with the Penrosian assumption of performance sustainability (Barney, 1991, Penrose 1959, Wernerfelt, 1984). It is however important to distinguish between the three types of RBV firms (Table 4).

**Table4: Three RBV Firms' Profiles**

| RBV Firms' Behavior | Market Context | Consequences |
|---|---|---|
| **Wallflowers** | No match with internal resource collection | Null or negative performance; endangered survival |
| **Convenience Marriage** | Monopoly of RBV firms | Low performance |
| **Soul Mates** | Hyper-competition with a lot of IO firms, with high gains expectancy. | Above IO firms performance |

In the majority of the performed simulations, some RBV firms do not find the adequate market: since they do not possess the relevant resources to enter one or another of the available markets, they keep out of the picture and do not perform at all. We call them "Wallflowers", as they look at other firms without participating in the economic game. If new markets appeared or entry conditions change, new opportunities could open for them where they could even over-perform as leaders. Since our market structure is static, those firms keep their wallflower status forever. Other RBV firms realize an average, if not a low performance, since the market that they chose because of its relevance for their internal competencies and on which they are installed is not very profitable, even if it allows them to survive. Their position on this market is like a "Convenience marriage" that does not provide them with the success they could hope but is preferable to be a "wallflower". At last, the highly performing RBV firms combine relevant resources for a given market to the luck that this market is very profitable. They consolidate their position on this hyper competitive market and perform systematically better than their IO competitors, which enter and leave the same market on the basis of, respectively, a high or low expected performance. We call these hyper-reactive firms Soul Mates.

We mainly explain the significant performance differences between IO and RBV firms by the impact on performance of the different analysis levels of the strategic model and by the market competitive structure. First, concerning the level of analysis, RBV and IO processes are very different. A firm adopting an IO strategy will necessarily find a market. Since market structure is the most important point for those firms (Bain, 1951, 1968; Mason, 1939, 1957; Porter, 1980a), it is highly likely that a market with the adequate structure exists. For RBV firms, on the contrary, finding a market on which their resource profile (Chatterjee & Wernerfelt, 1991) will be the vector of competitive advantage is more difficult. However, once such a market is fond, the firm is installed and stable. RBV firms are thus more dependent on the configuration of markets than IO firms.

Concerning the market structure, since IO firms can find rapidly a market corresponding to their strategic model, they reach also rapidly a high performance, while RBV firms are trying to sort themselves out. However, once RBV firms reach a profitable market position, the competitive environment and the equilibria between actors change (Van de Ven & Poole, 1995). Market structures evolve, forcing IO firms to adapt and seek other performance sources. At this point, RBV firms are installed and reach a higher performance. IO firms have to adapt constantly from market to market, while RBV firms also have to adapt but on a mastered market where the learning effect intervenes. IO firms obtain regularly over-performances when they move on a new market, but this over-performance is not sustainable and must be renewed. RBV firms, for their part, reach weaker but more regular performances. At the end of the simulation, after 200 periods, the sustainability of RBV firms seems to pay more than the feats of IO firms.

## 7. CONCLUSION

The goal of this study is to determine which strategic model, either IO or RBV, allows firms to reach the highest performance on a competitive market. Contrasting with classical studies that mobilize analyses as VARCOMP, we deploy a multi-agent system simulating the behavior of firms adopting RBV or IO strategic models. This research shows that, in a market with perfect competition composed of 200 firms and 20 markets, firms that have adopted the IO strategic model access more quickly to a very high performance. However, after a large number of periods, the highest performing firms are generally RBV firms.

In our view, the volatility in the results obtained in previous studies, summarized table 1, and the high level of error regularly obtained, are partly due to the versatility of RBV firms. Indeed, there is a very large variance in performance generated by RBV firms according to the type of firm (table 4). Thus, the results obtained in the precedent studies are based on the analysis of performance variance. In our view, the very different results obtained must be linked with the proportions of RBV firm of one kind or another in the population studied.

However, it is necessary to take these results for what they are. Indeed, the data used for our analysis are simulated data based on an algorithm that we developed based on theory. This research is therefore limited to the theoretical world, and this is fundamental. The pitfall of this type of research is to indulge the temptation to transpose early results in purely theoretical empirical recommendations. This transposition is neither founded, nor desirable, and in no cases our approach. Our objective was to determine the strategic model for the firm to generate more performance in a competitive market. This approach is based on a set of theoretical propositions that lead us to conclude that firms adopting the RBV strategic model access, on average, over a long period, and in the specific conditions of our simulations, to a significantly higher performance than that generated by firms IO.

**Appendix 1: Synthesis of the results at the end of the 20 cycles, for the 1008 simulations**

| | Nb. IO in best 10 | Nb. RBV in best 10 | Perf. 1st IO | Perf. 1st RBV | Av. Perf. 5 Best IO | Av. Perf. 5 Best RBV | Av. Perf. 10 Best IO | Av. Perf. 10 Best RBV | Av. Perf. all IO | Av. Perf. all RBV | Rel diff between best IO and best RBV[a] | Rel diff between av 5 best IO and av 5 best RBV[b] | Rel diff between av 10 best IO and av 10 best RBV[c] | Rel diff av all IO and av all RBV[d] |
|---|---|---|---|---|---|---|---|---|---|---|---|---|---|---|
| **Average** | 5,55 | 4,45 | 3447,23 | 3528,57 | 3259,60 | 3220,39 | 3146,82 | 3011,54 | 2511,61 | 1497,58 | -0,01 | 0,03 | 0,06 | 0,82 |
| **St. dev.** | 2,71 | 2,71 | 253,39 | 458,88 | 205,95 | 318,48 | 193,73 | 320,26 | 203,82 | 319,37 | 0,16 | 0,16 | 0,20 | 1,05 |
| **Variance** | 7,36 | 7,36 | 64204,78 | 210568,59 | 42413,81 | 101429,62 | 37530,87 | 102567,34 | 41541,46 | 101994,28 | 0,03 | 0,03 | 0,04 | 1,09 |
| **Median** | 6,00 | 4,00 | 3444,99 | 3483,70 | 3265,81 | 3234,13 | 3156,50 | 3052,31 | 2526,48 | 1512,14 | -0,02 | 0,00 | 0,03 | 0,66 |
| **Maxima** | 10,00 | 10,00 | 4152,83 | 4962,37 | 3873,67 | 4040,90 | 3673,89 | 3715,42 | 2999,23 | 2340,96 | 1,42 | 1,98 | 2,37 | 19,42 |
| **Minima** | 0,00 | 0,00 | 2673,57 | 1463,60 | 2595,96 | 1178,91 | 2533,42 | 1002,85 | 1796,39 | 127,59 | -0,37 | -0,28 | -0,24 | -0,05 |
| | | | | | | | | | **Nb of cases where IO > RBV** | | 443,00 | 511,00 | 605,00 | 1007,00 |
| | | | | | | | | | **Nb of cases where IO < RBV** | | 565,00 | 497,00 | 403,00 | 1,00 |
| | | | | | | | | | **% of cases where IO > RBV** | | 43,95% | 50,69% | 60,02% | 99,90% |
| | | | | | | | | | **% of cases where IO < RBV** | | 56,05% | 49,31% | 39,98% | 0,10% |

Where :

**a** : Relative difference between the best performing IO and RBV firms : (Best IO – Best RBV)/Best RBV
**b** : Relative difference between the average performance of the 5 best performing IO and RBV firms: (Best5IO – Best5RBV) / Best5RBV
**c** : Relative difference between the average performance of 10 best performing IO and RBV firms : (Best10IO – Best10RBV) / Best10RBV
**d** : Relative difference between the average performance of all remaining IO and RBV firms : (AvIO – AvRBV) / AvRB

## Appendix 2: Synthesis of the results at the end of the 200 cycles, for the 1008 simulations

| | Nb. IO in best 10 | Nb. RBV in best 10 | Perf. 1st IO | Perf. 1st RBV | Av. Perf. 5 Best IO | Av. Perf. 5 Best RBV | Av. Perf. 10 Best IO | Av. Perf. 10 Best RBV | Av. Perf. all IO | Av. Perf. all RBV | Rel diff between best IO and best RBV[a] | Rel diff between av 5 best IO and av 5 best RBV[b] | Rel diff between av 10 best IO and av 10 best RBV[c] | Rel diff av all IO and av all RBV[d] |
|---|---|---|---|---|---|---|---|---|---|---|---|---|---|---|
| **Average** | 3,93 | 6,07 | 5582,62 | 6654,33 | 5432,62 | 5985,43 | 5339,84 | 5585,26 | 4688,70 | 3055,44 | -0,14 | -0,08 | -0,02 | 0,66 |
| **St. dev.** | 3,27 | 3,27 | 294,05 | 937,24 | 287,43 | 632,86 | 292,25 | 616,07 | 344,52 | 615,50 | 0,17 | 0,17 | 0,20 | 0,89 |
| **Variance** | 10,68 | 10,68 | 86465,27 | 878420,96 | 82618,84 | 400512,42 | 85412,89 | 379540,78 | 118696,02 | 378843,46 | 0,03 | 0,03 | 0,04 | 0,80 |
| **Median** | 4,00 | 6,00 | 5597,86 | 6587,88 | 5443,05 | 6038,91 | 5350,85 | 5668,28 | 4714,08 | 3118,80 | -0,15 | -0,10 | -0,05 | 0,51 |
| **Maxima** | 10,00 | 10,00 | 6432,88 | 8490,24 | 6218,90 | 7566,40 | 6098,50 | 6812,13 | 5565,26 | 4546,53 | 2,08 | 2,35 | 2,60 | 14,25 |
| **Minima** | 0,00 | 0,00 | 4535,61 | 1827,73 | 4433,78 | 1672,18 | 4298,64 | 1560,73 | 3546,94 | 311,36 | -0,42 | -0,36 | -0,32 | -0,16 |
| | | | | | | | | | | **Nb of cases where IO > RBV** | 143,00 | 195,00 | 331,00 | 994,00 |
| | | | | | | | | | | **Nb of cases where IO < RBV** | 865,00 | 813,00 | 677,00 | 14,00 |
| | | | | | | | | | | **% of cases where IO > RBV** | 14,19 % | 19,35 % | 32,84 % | 98,61 % |
| | | | | | | | | | | **% of cases where IO < RBV** | 85,81 % | 80,65 % | 67,16 % | 1,39 % |

Where :
**a** : Relative difference between the best performing IO and RBV firms : (Best IO – Best RBV)/Best RBV
**b** : Relative difference between the average performance of the 5 best performing IO and RBV firms: (Best5IO – Best5RBV) / Best5RBV
**c** : Relative difference between the average performance of 10 best performing IO and RBV firms : (Best10IO – Best10RBV) / Best10RBV
**d** : Relative difference between the average performance of all remaining IO and RBV firms : (AvIO – AvRBV) / AvRB